\documentclass[journal,twoside,web]{ieeecolor}
\usepackage{etoolbox}
\pdfoutput=1%For ARXIV
\UseRawInputEncoding
\makeatletter
\@ifundefined{color@begingroup}%
{\let\color@begingroup\relax
\let\color@endgroup\relax}{}%
\def\fix@ieeecolor@hbox#1{%
\hbox{\color@begingroup#1\color@endgroup}}
\patchcmd\@makecaption{\hbox}{\fix@ieeecolor@hbox}{}{\FAILED}
\patchcmd\@makecaption{\hbox}{\fix@ieeecolor@hbox}{}{\FAILED}
\usepackage{generic}
\usepackage{cite}
\usepackage{amsmath,amssymb,amsfonts}
\usepackage{algorithmic}
\usepackage{graphicx}
\usepackage{textcomp}
\usepackage{multirow}
\usepackage{array}
\usepackage{makecell}
\usepackage{color}
\usepackage{soul}
\usepackage[export]{adjustbox} % 添加 adjustbox 宏包
\usepackage{bm}
\usepackage{tabularx}
\usepackage{caption}
\usepackage{booktabs} 
\makeatletter
\let\NAT@parse\undefined
\makeatother
\usepackage{hyperref}
\hypersetup{colorlinks=true,
           linkcolor=blue,
           anchorcolor=blue,
           citecolor=green,
           urlcolor=blue}
\def\BibTeX{{\rm B\kern-.05em{\sc i\kern-.025em b}\kern-.08em
    T\kern-.1667em\lower.7ex\hbox{E}\kern-.125emX}}
\begin{document}
\title{UWAFA-GAN: Ultra-Wide-Angle Fluorescein Angiography Transformation via Multi-scale Generation and Registration Enhancement}
\author{Ruiquan Ge, \IEEEmembership{Member, IEEE}, Zhaojie Fang, Pengxue Wei, Zhanghao Chen, Hongyang Jiang, Ahmed Elazab, Wangting Li, Xiang Wan, Shaochong Zhang, Changmiao Wang
\thanks{This work was supported in part by the Zhejiang Provincial Natural Science Foundation of China (No.2022C03043, LY21F020017), National Natural Science Foundation of China (No.U20A20386, U22A2033, 61702146),  Chinese Key-Area Research and Development Program of Guangdong Province (2020B0101350001), GuangDong Basic and Applied Basic Research Foundation (No. 2022A1515110570), Innovation teams of youth innovation in science and technology of high education institutions of Shandong province (No. 2021KJ088), the Shenzhen Science and Technology Program (JCYJ20220818103001002), and the Guangdong Provincial Key Laboratory of Big Data Computing, The Chinese University of Hong Kong, Shenzhen. (Corresponding author: Changmiao Wang, Shaochong Zhang).}
\thanks{R. Ge, Z. Fang, and Z. Chen are with Hangzhou Dianzi University, Hangzhou, 310018, China (e-mail:(gespring,fangzhaojie, czh15857582366)@hdu.edu.cn).}
\thanks{H. Jiang is with Southern University of Science and Technology, Shenzhen, 518055, China (e-mail: jianghy3@sustech.edu.cn).}
\thanks{A. Elazab is with the School of Biomedical Engineering, Shenzhen University, Shenzhen, China and Computer Science Department, Misr Higher Institute for  Commerce and Computers, Mansoura, Egypt (e-mail:ahmed.elazab@yahoo.com).}
\thanks{P. Wei, W. Li and S. Zhang are with Shenzhen Eye Hospital, Jinan University, Shenzhen 518040, China (e-mail:wei542172846@outlook.com; liwangting@hotmail.com; zhangshaochong@gzzoc.com).}
\thanks{X. Wan and C. Wang are with Shenzhen Research Institute of Big Data, Shenzhen, 518172, China (e-mail:wanxiang@sribd.cn; cmwangalbert@gmail.com).}
}

\maketitle

\begin{abstract}
Fundus photography, in combination with the ultra-wide-angle fundus (UWF) techniques, becomes an indispensable diagnostic tool in clinical settings by offering a more comprehensive view of the retina. Nonetheless, UWF fluorescein angiography (UWF-FA) necessitates the administration of a fluorescent dye via injection into the patient's hand or elbow unlike UWF scanning laser ophthalmoscopy (UWF-SLO). To mitigate potential adverse effects associated with injections, researchers have proposed the development of cross-modality medical image generation algorithms capable of converting UWF-SLO images into their UWF-FA counterparts. Current image generation techniques applied to fundus photography encounter difficulties in producing high-resolution retinal images, particularly in capturing minute vascular lesions. To address these issues, we introduce a novel conditional generative adversarial network (UWAFA-GAN) to synthesize UWF-FA from UWF-SLO. This approach employs multi-scale generators and an attention transmit module to efficiently extract both global structures and local lesions. Additionally, to counteract the image blurriness issue that arises from training with misaligned data, a registration module is integrated within this framework. Our method performs non-trivially on inception scores and details generation. Clinical user studies further indicate that the UWF-FA images generated by UWAFA-GAN are clinically comparable to authentic images in terms of diagnostic reliability. Empirical evaluations on our proprietary UWF image datasets elucidate that UWAFA-GAN outperforms extant methodologies. The code is accessible at \href{https://github.com/Tinysqua/UWAFA-GAN}{https://github.com/Tinysqua/UWAFA-GAN}.
\end{abstract}

\begin{IEEEkeywords}
Fluorescein angiography, Cross-modality image generation, Ultra-wide-angle fundus imaging, Conditional generative adversarial network, Supervised learning.
\end{IEEEkeywords}

\section{Introduction}
\label{sec:introduction}
\IEEEPARstart{R}{etinal} diseases are prevalent and serious eye conditions in today's populations. Fluorescein angiography (FA) and fundus photograph (FP) are complementary imaging techniques widely utilized for the detection and diagnosis of these conditions. FA enables the imaging of vascular structures and observation of blood vessels but requires the injection of fluorescent dyes into the patient's body via the elbow veins or hands. Such injection can potentially lead to severe allergic reactions, pain in the arm\cite{tavakkoli2020novel}, and retinal vascular diseases\cite{kawai2023clinically}. Conversely, FP, another modal revealing retinal vascular structures, does not require dye injection but lacks the clarity provided by FA imaging. In recent years, the advent of ultra-wide-angle fundus (UWF) imaging systems has revolutionized conventional FA and FP, yielding UWF-FA and UWF-scanning laser ophthalmoscopy (UWF-SLO), respectively. The UWF imaging system captures a broader perspective in a more time-efficient manner, which is advantageous for screening and monitoring the progression of retinal vascular diseases\cite{wang2021automated,ashraf2020diabetic}. Nonetheless, the adverse impacts associated with FA persist. Therefore, exploring effective techniques to convert UWF-SLO to UWF-FFA holds significant potential value.

Cross-modal medical image synthesis makes this conversion possible.  In the realm of cross-modal medical image generation, generative adversarial networks (GANs) \cite{goodfellow2020generative} and their variants \cite{chen2016infogan, mehralian2018rdcgan} have brought breakthroughs in multi-modal medical image synthesis, proving vastly superior to generation-focused models using convolution neural network (CNN) that are prone to producing blurry outcomes, particularly when tasked with creating more realistic images. Park et al. \cite{park2020m} presented a novel conditional GAN (cGAN) termed M-GAN for precise retinal vascular segmentation. M-GAN includes an M-generator with deep residual blocks and an M-discriminator, which work together to enable robust segmentation and quick adversarial training. Kamran et al. \cite{kamran2021vtgan} developed a VTGAN, a semi-supervised conditional GAN, that can create retinal vascular architecture from FA while discriminating between healthy and sick retina. By combining multi-scale coarse and fine generators, VTGAN could collect features across scales and yield more lifelike images. However, considering the UWF scope's exceptionally high resolution, the aforementioned techniques often required substantial memory and proved inefficient for training. Furthermore, they tended to yield suboptimal results in generating detailed vascular and lesion imagery within specific localized areas.

Besides GAN, diffusion models \cite{nichol2021improved} continue to produce remarkable results. These models improved and gradually expanded from random image generation to conditional controlled generation, thereby being subject to more precise constraints, such as text \cite{nichol2021glide, DBLP:conf/icml/RadfordKHRGASAM21}, class labels \cite{ho2021classifierfree, yang2022diffusion}, and images \cite{saharia2022palette}. In the realm of medical images, SynDiff \cite{Syndiff} amalgamates adversarial networks with diffusion models to efficiently perform MRI-CT translations on unpaired data. Despite these advancements, challenges remain, including the synthesis of lower-resolution images, loss of modal data during fusion, and the discriminator's limitation to patch-level information \cite{isola2017image}. Nevertheless, a notable limitation is that these networks, even those as robust as diffusion models, are optimized primarily for global features and not for the detection of minute lesions\cite{akram2014detection}. Moreover, the efficacy of diffusion models diminishes with smaller datasets. These shortcomings render the application of diffusion-based models to UWF image generation.

Another challenge in this field is the misalignment between paired images, which often occurs due to involuntary eye movements during the acquisition of UWF-SLO and UWF-FA images. Efficiently training models on such unpaired data is a formidable task. Prior research attempted to mitigate the issue by utilizing partially registered paired images. For instance, Patrini et al. proposed a method called 'loss-correction'  \cite{patrini2017making} to model the deformation field before conducting resampling, thereby achieving a training effect equivalent to that of clean labels on noisy labels. Subsequently, Goldberger et al. \cite{8363518} conceptualized clean labels as a distribution within the latent space and developed a congruent model and loss function to identify a similar distribution within the space. Kong et al. \cite{kong2021breaking} first used the U-Net model to fit the deformation field and introduced a smoothness loss to ensure fluidity within the deformation field. Even though their work allows us to complete registration and conversion into two separate steps and models, we believe that both are intrinsic to CNNs, and the fact that they can be optimized simultaneously prompts us to consider integrating the registration process with GAN.

Considering the existing challenges and motivated by pix2pixHD \cite{wang2018high}, we introduce the Ultra-Wide-Angle Fluorescein Angiography GAN (UWAFA-GAN) with multi-scale generators to produce UWF-FA from UWF-SLO, capturing minute vascular lesions. Additionally, we integrate a registration module to mitigate the impact of image misalignment noise on the training process and demonstrate its efficacy in enhancing model training across different levels of noise. Equipped with the registration enhancement module and multi-scale generators, our model synthesizes high-resolution, rectangular images and improves the capability to capture minute vascular lesion areas. We further employ various weighted losses at different data scales to ensure optimal training of the model.

In terms of evaluation metrics, we employ Fréchet inception Distance (FID)\cite{binkowski2018demystifying}, Inception Score (IS) \cite{chong2020effectively}, Multi-Scale  Structural Similarity Index Measure (MS-SSIM), and Peak Signal-to-Noise Ratio (PSNR) \cite{sara2019image} to assess the quality of images generated by the model. We compare our proposed architecture with the state-of-the-art synthesis models in fundus photography for qualitative assessment and visual comparison. To underscore the effectiveness of the registration module in our framework, we conducted an ablation study where we manually introduced several levels of noise to the paired images. Furthermore, we implemented a clinical user study involving three ophthalmologists of varying levels of expertise. The fact that a significant number of generated images were indistinguishable from real ones, as evidenced by the misjudgment of the ophthalmologists, underlines the high clinical utility of our method. We posit that our model has the potential to be universally applied to numerous downstream tasks, such as clinical diagnosis and detection, thereby making a substantial contribution to the field of medicine.

\begin{figure*}[!tbp]
\centering
\includegraphics[width=7.16in, height=2.5in]{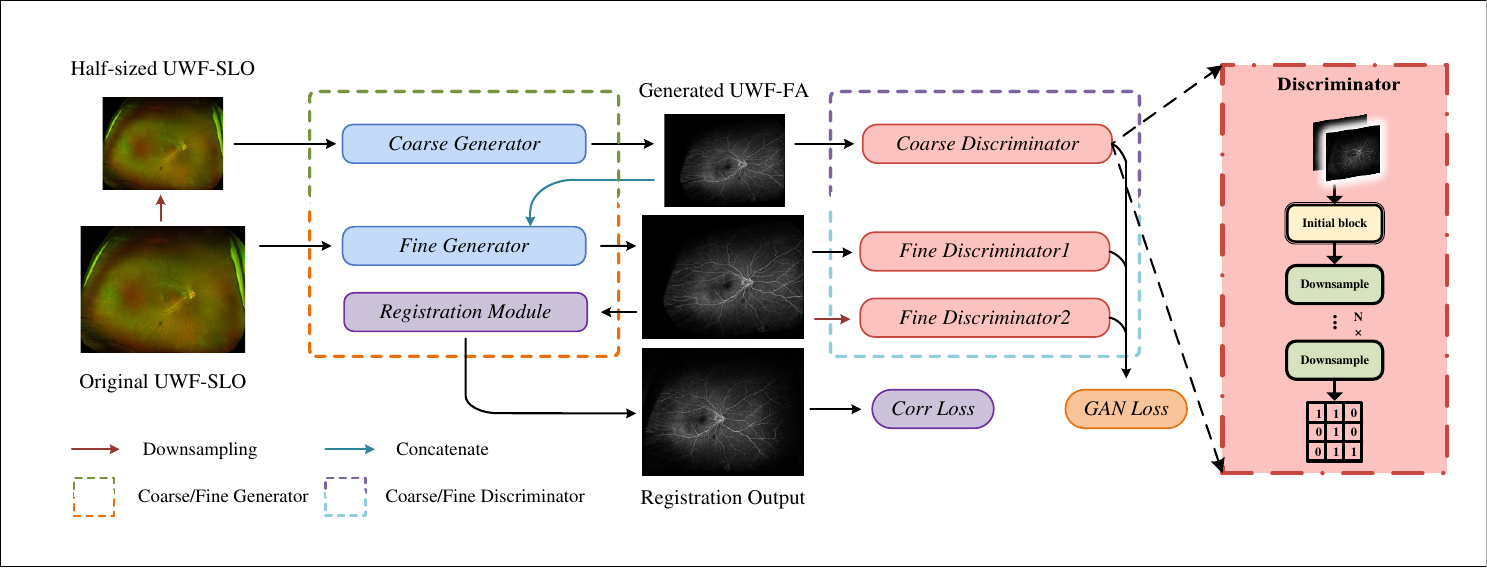}
% \centerline{\includegraphics[width=\columnwidth]{LaTeX/Pic/TMI_overall_architecture.pdf}}
\caption{The comprehensive architecture of our proposed UWAFA-GAN model.}% The architecture consists of two principal components: the Coarse Generator and the Fine Generator. The multi-scale discriminator is represented by two discriminators, Fine Discriminator1 and Fine Discriminator2, which process the outputs from the generators and compute the GAN loss. The registration module takes label $y$ and the output from the Fine Generator (denoted as $G_F$) to calculate the Corr loss.}
\label{architecture}
\end{figure*}

Our main contributions are summarized as follows:

1) We introduce, to the best of our knowledge, the first study to synthesize UWF-FA from UWF-SLO, thereby surmounting the inherent limitations of invasive UWF-FA imaging.

2) The proposed UWAFA-GAN leverages multi-scale generators and enhancement registration modules to produce high-fidelity images, capturing fine details like capillary perfusion zones. Furthermore, we have substantiated the superiority of UWAFA-GAN over existing SOTA models through inception metrics and visual comparisons.

% 3) We evaluate the performance of UWAFA-GAN using our self-collected datasets, implementing an efficacious preprocessing technique for image sharpening and registration. This improves the clarity of vascular regions, effectively addressing the issue of misalignment and facilitating a superior distinction between blood vessels and the image background.

3) We employ an efficacious image sharpening to distinguish blood vessels from the background and a registration module to effectively train on misaligned data, which have been evaluated by an ablation study.

4) Through rigorous clinical studies and comprehensive downstream task analyses, we have demonstrated the significant clinical utility and the beneficial impact on downstream clinical tasks of the proposed UWAFA-GAN. 

 % We have changed the details of the module, evaluation, image processing and added more experiments with new datasets.

\begin{figure}[!tbp]
\centerline{\includegraphics[width=\columnwidth]{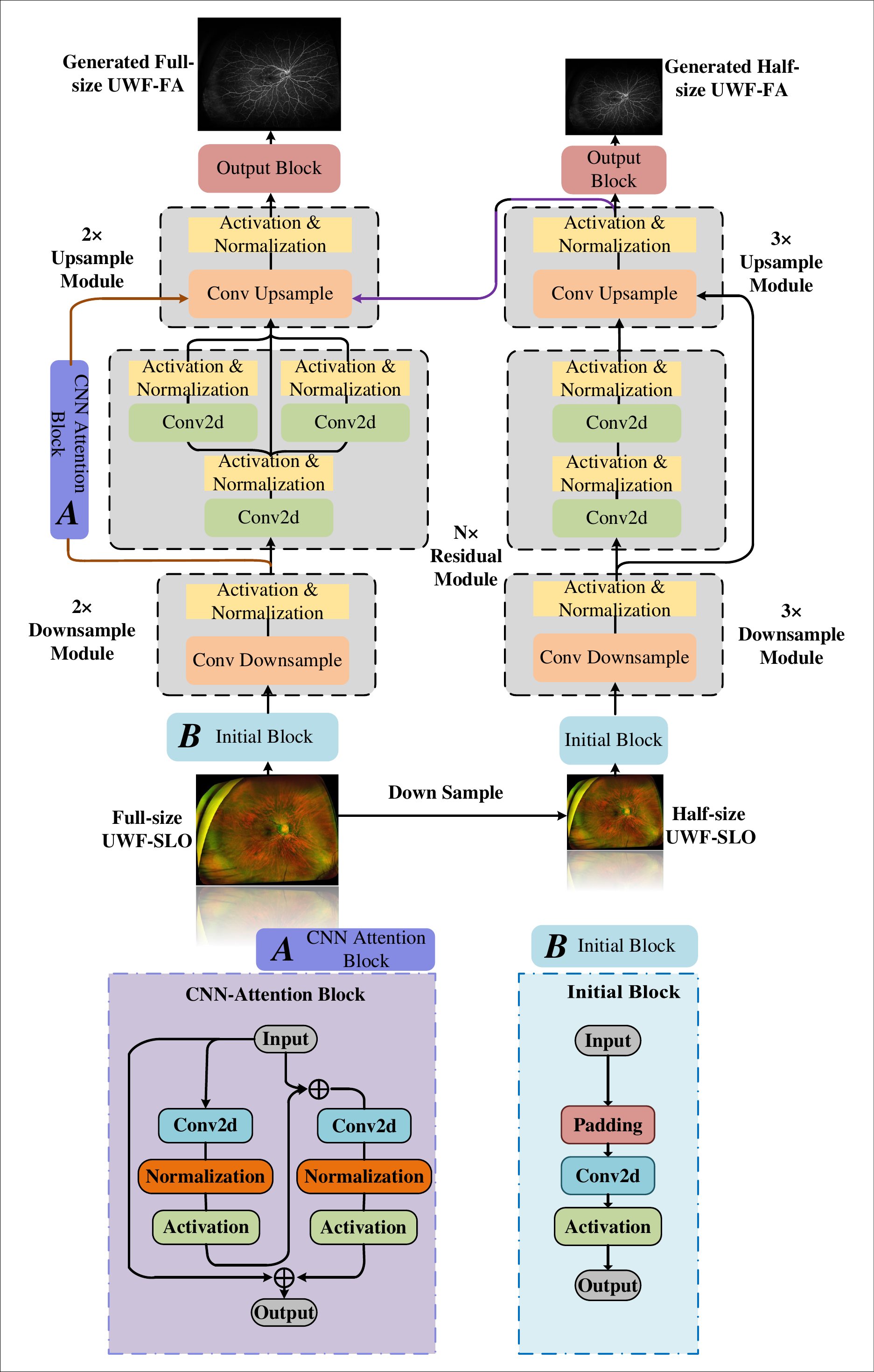}}
\caption{Generators' architectures of the UWAFA-GAN; The fine-level generator (left) and the coarse-level generator (right). The CNN Attention Block and Initial Block are marked with A and B.}
\label{Overall}
\end{figure}

\section{Methods}
We propose a supervised GAN that utilizes images as a condition to convert UWF-SLO into UWF-FA. This transformation ensures the clarity of vascular textures and specific lesions in the generated images. The complete architecture is delineated in Fig. \ref{architecture}. To accomplish this conversion, we introduce a module that amalgamates the fine-coarse level generator (\ref{level}). Moreover, we employ the registration module to shield the generator from potential data noise, thereby facilitating enhanced image generation (\ref{register}). Detailed descriptions of the generator blocks are provided, particularly those about the upscaling and downscaling processes (\ref{detail}). An additional element, the attention transmission block, has been incorporated into the skip connection (\ref{attention}). Ultimately, we delve into the specifics of the multi-scale discriminator (\ref{multi}) and the individual components of the loss function, along with their corresponding weights (\ref{loss_func}).

\subsection{Fine-Coarse Level Generator}\label{level}
Similar to other medical images, UWF-FA captures a plethora of local features, such as minute lesions and capillary blood vessels, as well as various global features, including the overall structure of the eye and elongated veins. Consequently, deploying a standard generator in cGAN to fabricate convincing images can be challenging. To address this issue, we propose a structure comprising two levels of generators. The coarse generator ($Gen_C$) extracts global information, subsequently yielding results based on this global context, and the fine generator ($Gen_F$) is dedicated to local information extraction. This dual approach ensures a balanced extraction and application of both global and local information (Fig. \ref{architecture}). We supply both the normal-sized UWF-SLO and its half-sized downsampled counterpart to $Gen_F$ and $Gen_C$, respectively. The input to the $Gen_C$ passes through three layers of the downsample module, six layers of the residual module, and three layers of the upsample module, ultimately yielding a feature map. This map is subsequently fed into the upsample module of the $Gen_F$, which assists in the generation of a higher-resolution image. The $Gen_F$ utilizes the same downsample, residual, and upsample modules as the $Gen_C$ but additionally includes a CNN attention block. The relevance and mechanism of the attention block will be discussed in further detail in Section \ref{attention}. The UWF-FA images generated by $Gen_F$ will be evaluated in subsequent experiments.

\subsection{Registration Module}\label{register}
A notable challenge previously identified is the misalignment observed between UWF-SLO and UWF-FA images. In particular, it is unable to guarantee precise congruence of retinal blood vessels across UWF-SLO and UWF-FA images, resulting in poor training outcomes. Therefore, we denote the misaligned UWF-SLO and UWF-FA pairs as ${x}$ and $y$ respectively. We introduce $\tilde{y}$ to represent the hypothetical UWF-FA that would be perfectly aligned with the corresponding UWF-SLO. Recognizing that $\tilde{y}$ represents a latent and idealized distribution, our goal is for the output of the generation process, \emph{$G_F(x)$}, to approximate the distribution of $\tilde{y}$ as closely as possible. Accordingly, we construct the following objective function to promote alignment correction:
\begin{equation}
    \min_{G_F}\mathcal{L}_{Corr}(T \odot G_F(x), y),
\end{equation}
where \emph{T} signifies a transition associated with noise. Patrini et al. \cite{patrini2017making} have shown that the distribution of the generative model's output, \emph{$G_F(x)$}, will converge to the latent true distribution $\tilde{y}$ when \emph{T} closely models the underlying noise distribution. Consequently, we aim to accurately estimate the transition \emph{T}. To calculate \emph{T}, we employ a single 5-stage UNet \cite{10.1007/978-3-319-24574-4_28}, designated as $\psi$, which serves as the registration module. This module is composed of four encoding and four decoding blocks. Each encoding block includes a convolution layer with a stride of 2 and one ResNet block \cite{he2015deep}. The decoding blocks are structurally analogous to the encoding blocks, except that the convolution layers are replaced by transposed convolution layers. The registration module takes $y$ and \emph{$G_F(x)$} as inputs and yields the transition \emph{T} as its output. As illustrated in Fig. \ref{architecture}, the registration module receives input exclusively from the fine generator, rather than participating in the generation and training processes of the coarse generator. This design choice is intentional; the coarse generator's role is to furnish the fine generator with coarse-grained information via intermediate vectors. Within the latent space, these vectors serve a pivotal function, conveying critical information without necessitating precise geometric alignment.

Then, we introduce the following loss function to measure the disparity between \emph{T} and the target distribution $y$: 
\begin{equation}
    \mathcal{L}_{Corr} = \mathcal{L}_{VGG}(\psi(G_F(x), y) \odot G_F(x), y),
\end{equation}
where $\mathcal{L}_{VGG}$ represents the perceptual loss  \cite{yang2018low}, see \ref{loss_func}.

\subsection{Details of Generator Blocks}\label{detail}
In the process of determining the optimal number of downsampling layers for our generators, we conducted a thorough evaluation of various configurations. Ultimately, we chose to implement three downsampling layers in the $Gen_C$ and two downsampling layers in the $Gen_F$. Our generators share several other architectural components, including the Initial block, Downsample block, Upsample block, Residual block, and Attention Transmit block (Fig. \ref{architecture}).

The generator's Initial block features a Leaky ReLU, Reflection padding, and a 2D convolution (stride 1, kernel size 1). The Downsample block includes a 2D convolution, instance normalization, and Leaky ReLU, while the Upsample block has a Transposed 2D convolution, instance normalization, and Leaky ReLU. The 2D convolution in both Downsample and Upsample blocks uses a 3 kernel size and a stride of 2.

\subsection{Attention Transmit Block}\label{attention}
In traditional UNet architectures, skip connections are employed to directly transmit the output from each encoding block to the corresponding decoding block. This approach aims to preserve fine-grained details that might otherwise be lost during the encoding phase. However, when generating UWF-FA images from UWF-SLO data, the density of relevant information is not always high. The generation process involves components such as: (a) the patient's eye socket and eyelashes, (b) numerous non-pathological regions, and (c) sparse retinal vessels. Given these conditions, it may not be advisable to forward all feature maps from the encoding phase directly to the decoding phase without selective filtering.

To address this, we introduce an Attention Transmit block that processes the feature maps before they reach the decoder. This block enables the decoder to selectively focus on and incorporate salient detailed information, which is particularly valuable in medical image generation where precision is crucial. The architecture of the Attention Transmit block is designed to potentially include mechanisms such as multi-head attention, cross-attention, or a series of convolutional layers. The configuration of this block is shown in Fig. \ref{architecture}.

\subsection{Multi-Scale Discriminator}\label{multi}
Our approach is influenced by the multi-scale discriminator used in the pix2pixHD model, which operates across different scales to differentiate between genuine and synthetic images, leveraging both global and local receptive fields. During the backpropagation process, the cumulative loss from the two levels of the discriminator is calculated, with each loss component appropriately weighted and then combined.

Let $Y_1$ represent the image outputted by the generator $Gen_X$, which is subsequently fed into the first discriminator, denoted as $D_{X1}$. Following this, we introduce $Y_{1/2}$, which is a downscaled version of $Y_1$ by a factor of two, and it is presented to the second discriminator, $D_{X2}$. In theory, one could construct an image pyramid $\{Y_1, Y_{1/2}, \ldots, Y_{1/2^{n-1}}\}$ and correspondingly employ a series of discriminators $\{D_{X1}, D_{X2}, \ldots, D_{Xn}\}$. Nonetheless, within our specific framework, we assign discriminators $D_{F1}$ and $D_{F2}$ to process the outputs of $Gen_F$, while the discriminator $D_C$ is utilized for the outputs of $Gen_C$.

\subsection{The Cost Functions}\label{loss_func}
Let's denote the fine and coarse generators as $G_F$ and $G_C$, respectively. A multi-scale discriminator, comprising $D_{F1}$ and $D_{F2}$, discriminates images from $G_F$, while a single-scale discriminator, $D_C$, discriminates images from $G_C$. We denote a pair of variables as \{($c_i, y_i$)\}, where $c$ embodies the distribution of the input for the entire model as a condition for generation (in our case, UWF-SLO), and $y$ represents the real distribution of the model's output (UWF-FA). Consequently, the objective function, given $c$, aims to maximize the loss of $D_{F1}$, $D_{F2}$, and $D_C$, while simultaneously minimizing the loss of $Gen_C$ and $Gen_F$, is expressed as follows:
\begin{equation}
\begin{split}
\label{B}
    \min_{G_F}\max_{D_{F1},D_{F2}}\sum_{k=1,2}\mathcal{L}_{cGAN}(G_F, D_{Fk}) \\
    + \min_{G_C}\max_{D_{C}}\mathcal{L}_{cGAN}(G_C, D_{C}),
\end{split}
\end{equation}
where $\mathcal{L}_{cGAN}$ is given by:
\begin{equation}
    \mathbb{E}_{(c,y)}[log(D(c,y))]+\mathbb{E}_c[log(1-D(G(c),c)].
\end{equation}
To enhance the evaluation of image quality beyond the conventional adversarial loss, we incorporate the feature mapping (FM) loss \cite{simonyan2015deep} into our loss measurement apparatus. This process begins by partitioning the layers within the discriminators and preserving their respective outputs. Subsequently, we denote the output from the $i^{th}$ partitioned layer in the discriminator as $D^{(i)}$. Given a set of tuples comprising the conditions for generation, the ground truth output, and the images synthesized by the generator, denoted as \{$c, y, G(c)$\}, we articulate the loss function as follows:
\begin{equation}
%\begin{split}
\setlength\abovedisplayskip{7pt}
\setlength\belowdisplayskip{7pt}
\label{C}
    \mathcal{L}_{FM}(G, D_k) = 
    \mathbb{E}_{(c,y)}\sum_{i=1}^M\frac{1}{N_i}[\left\|D^i_k(c,y)-D^i_k(c,G(c))\right\|_1],
%\end{split}
\end{equation}
where $M$ signifies the total count of partitioned layers within the discriminator, while $N_i$ indicates the number of elements contained within each layer, such as convolution blocks, normalization components, and activation functions. The minimization of this loss aims to diminish the discrepancy between the features extracted from the paired set \{$y, G(c)$\} by each discriminator layer, thereby augmenting the discriminator's capacity to differentiate between real and generated images. 

In addition to the aforementioned loss, our model employs perceptual loss rather than the conventional L1 loss. This perceptual loss is computed using a pre-trained VGG19 network, which serves to extract and analyze the features from the paired images \{$y, G(c)$\}. As outlined in \ref{register}, our primary goal is to mitigate misalignment issues between the condition $c$ and the ground truth $y$, guiding the generated image $G(c)$ closer to the target latent distribution $\tilde{y}$. Assuming the presence of a registration module denoted as $\psi$, the loss function related to this objective is expressed as follows:
\begin{equation}
\setlength\abovedisplayskip{7pt}
\setlength\belowdisplayskip{7pt}
%\begin{split}
\label{D}    
    \mathcal{L}_{VGG}(G, D_k) = \sum^N_{i=1}\frac{1}{M_i}[\left\|V^i(y)-V^i(\psi(G(c)))\right\|_1],
%\end{split}
\end{equation}
where $N$ represents the number of layers in the VGG-19 network. The term $\frac{1}{M_i}$ ensures averaging across each layer. We represent the $i^{th}$ layer of the VGG19 network with $V^i$. The formulation of the final cost function is established as:

\begin{equation}\label{E}
\begin{aligned}
        \min_{G_C}(\max_{D_{C1},D_{C2}}\sum_{k=1,2}\mathcal{L}_{cGAN}(G_C, D_{Ck})& \\  + \lambda_{FMC}\sum_{k=1,2}\mathcal{L}_{FM}(G_C, D_{Ck}) & \\
        + \lambda_{VGGC}\sum_{k=1,2}\mathcal{L}_{VGG}(G_C, D_{Ck})) &\\
    +  \min_{G_F}(\max_{D_F}\mathcal{L}_{cGAN}(G_F, D_{F}) &\\
    +  \lambda_{FMF}\mathcal{L}_{FM}(G_F, D_{F})\\ +  \lambda_{VGGF}\mathcal{L}_{VGG}(G_F, D_{F}))&,
\end{aligned}
\end{equation}
where $\lambda_{FMC},\lambda_{VGGC},\lambda_{FMF}$,  and $\lambda_{VGGF}$ indicate adjustable weight parameters.

\begin{table*}[!t]
\scriptsize
\centering
\caption{Comparison with the state-of-the-art methods utilizing
four distinct evaluation metrics on $Intra$ and $Combined$ datasets. The best and second-best performances are indicated in \textcolor{red}{red} and \textcolor{blue}{blue} colors, respectively.}
\label{tab1}
    \centering
    \begin{adjustbox}{width=0.9\textwidth, center}
        \begin{tabular}{cccccccccc}
        \toprule
            \multicolumn{2}{c}{\multirow{2}{*}{Method}} & \multicolumn{4}{c|}{$Intra$} & \multicolumn{4}{c}{$Combined$} \\
            \cline{3-10}
            \addlinespace
             ~ & ~ & FID $\downarrow$ & IS $\uparrow$ & MS-SSIM $\uparrow$ & PSNR $\uparrow$ & FID $\downarrow$ & IS $\uparrow$& MS-SSIM $\uparrow$& PSNR$\uparrow$  \\
        \midrule
        ~ & Pix2Pix\cite{isola2017image} & 76.744 & 0.514 & 0.477 & 19.149 & 69.392 & 0.554 & 0.506 & 20.11 \\ 
            ~ & Pix2PixHD\cite{wang2018high} & 38.054 & 1.246 & 0.701 & 26.231 & 30.338 & 1.039 & 0.801 & 26.144  \\ 
        ~ & Reg-GAN\cite{fang2023uwat} & 38.571 & 1.217 & 0.651 & 24.204 & 32.404 & 1.377 & 0.899 & 25.838  \\ 
        % \midrule
         ~ & ControlNet\cite{zhang2023adding} & \textcolor{blue}{27.953} & \textcolor{blue}{1.591} & \textcolor{blue}{0.839} & 28.938 & \textcolor{blue}{22.427} & \textcolor{blue}{1.498} & \textbf{\textcolor{red}{0.994}} & \textcolor{blue}{31.226}  \\ 
        ~ & Latent Diffusion\cite{rombach2022high} & 33.501 & 1.241 & 0.735 & \textcolor{blue}{28.959} & 27.429 & 1.388 & 0.902 & 29.102  \\ 
        \midrule
        ~ & \textbf{UWAFA-GAN (Ours)} & \textbf{\textcolor{red}{21.426}} & \textbf{\textcolor{red}{1.672}} & \textbf{\textcolor{red}{0.933}} & \textbf{\textcolor{red}{32.771}} & \textbf{\textcolor{red}{15.922}} & \textbf{\textcolor{red}{1.691}} & \textcolor{blue}{0.958} & \textbf{\textcolor{red}{33.908}} \\  
        \bottomrule
        \end{tabular}
    \end{adjustbox}
% \begin{table}[!h]
\end{table*}

\section{Experiments}

\subsection{Data Preparation and Processing}\label{preparation}

The dataset underpinning our experimental framework was procured through a collaborative effort with allied eye hospitals. To fully demonstrate the superiority of our method over existing methods, we constructed two datasets. Our method aims to generate intravenous UWF-FA through UWF-SLO by injecting contrast agents. The decision to construct two distinct datasets is predicated on the presence of two types of UWF-FA images: intravenous and oral. Despite both modalities providing analogous results regarding vascular structure and lesion depiction, intravenous UWF-FA is associated with a higher potential risk for patients. However, it delivers enhanced clarity in vascular imaging and facilitates more effective detection of certain rare lesions compared to the oral counterpart. From a clinical standpoint, the generation of high-quality intravenous UWF-FA images is crucial and is, therefore, the primary focus of our research. Nonetheless, the intravenous approach is marked by significant drawbacks such as elevated costs, prolonged acquisition times, and a limited patient cohort with corresponding lesions. Consequently, the number of paired UWF-SLO and intravenous UWF-FA images is relatively scarce within our dataset, comprising 164 pairs, which we refer to as the $Intra$ dataset. To demonstrate the scalability and generalizability of our proposed method, especially in the context of inception metric studies, we have also included paired UWF-SLO and orally administered UWF-FA images, amounting to a total of 140 pairs. This alternative mode of administration offers a less invasive and potentially safer means of obtaining UWF-FA images. The amalgamation of UWF-FA images from both intravenous and oral protocols yields a cumulative dataset of 304 pairs, referred to as the $Combined$ dataset. The comparative analysis of the $Intra$ and $Combined$ datasets is instrumental in establishing the superiority of our method. The results underscore our methodology's efficacy in addressing both clinically specialized tasks and technological benchmarks. We uniformly follow the same preprocessing steps on both datasets, including image sharpening, data augmentation, and splitting into train and test sets.

% Following a stringent quality assessment, our final dataset includes 303 paired images, each with a resolution of 1112 × 1448 pixels. We designated 70\% of these image pairs for the training set and allocated the remaining 30\% for validation. It is noteworthy that the relatively modest size of our dataset, in comparison to other medical imaging studies, is a consequence of the time-intensive procedure required for the administration of fluorescent dyes and the comparatively infrequent incidence of patients subjected to this diagnostic test. Despite the limited dataset size, the comprehensive nature of the UWF images can yield significant insights and a wealth of information. Moreover, to circumvent potential issues stemming from the limited diversity of the dataset and enhance the robustness of our model, we implemented appropriate data augmentation techniques.

To improve image visual clarity, we applied the Contrast Limited Adaptive Histogram Equalization (CLAHE) technique, which enhances image sharpness through localized contrast augmentation. Unlike global histogram equalization, CLAHE operates on discrete sections, or tiles, of an image, optimizing the contrast within each tile based on a predefined histogram distribution. The method then employs bilinear interpolation to seamlessly merge these adjusted tiles, ensuring smooth transitions across tile boundaries. This localized contrast enhancement is particularly effective in heightening the distinction between blood vessels and the surrounding tissue, a process known as Image Sharpening (ImS).

% This technique aids in making local features more discernible, particularly in regions that are overly dark or light. Post histogram equalization, the preprocessed images exhibit a heightened contrast differentiation between the coloration of the blood vessels and the surrounding background, thereby enriching the representation of the vascular structure and facilitating superior discrimination between individual vessels.

Data augmentation is further adopted to enrich our experimental image dataset.  The paired images were first employed with a random flipping, then a random cropping. The resolution for random cropping was set at 1088 × 832 pixels, a size selected to consistently encompass the central region of the original images, compared to the original 1112×1448 pixels, which is typically abundant in features associated with ocular blood vessels. From each original image pair, we extracted 40 cropped pairings, culminating in an expanded dataset comprising 6560 images for $Intra$ dataset and  12160 for $Combined$ one. Given that the distribution of lesions within the images is both scattered and stochastic, two sub-images cropped from the same original may exhibit markedly distinct characteristics, capturing a variety of minute lesions and retinal structures.

% To rigorously evaluate the efficacy of the image registration module, we deliberately introduced noise during the training phase. For each newly generated batch, the images were modified through a series of transformations: rotations within a range of -20 to 20 degrees, horizontal and vertical translations of roughly 17 and 22 pixels, respectively, and a zoom factor with a scaling ratio of approximately 1±0.02.

Finally, we designated 70\% of the two datasets for the training set and allocated the remaining 30\% for evaluation. In the upcoming experiment, the inception metrics study will be conducted on both the $Intra$ and $Combined$ datasets to investigate the performance of methods on different sizes datasets. The ablation study, on the other hand, will only be conducted on the $Intra$ dataset.

\subsection{Implementation}
\subsubsection{Comparative Methods}
We conducted a comparative analysis of the proposed method against four state-of-the-art conditional generation models: Pix2Pix, Pix2PixHD, Reg-GAN, Latent diffusion, Controlnet. The efficacy of these models was determined by evaluating the quality of the UWF-FA images they generated. Pix2Pix, which leverages GAN architecture for conditional image generation, served as our baseline model. Pix2PixHD, an extension of Pix2Pix, introduced significant enhancements, particularly in generating high-resolution images.

Reg-GAN \cite{kong2021breaking} incorporated a registration network to address the challenge of image translation in misaligned datasets. The latent diffusion model \cite{rombach2022high} represented a new approach to image generation. Expanding on the foundational architecture of the diffusion model, latent diffusion involves the compression of images into a latent space, where conditional images are concatenated with target images. This combined representation in the latent space enables the denoising mechanism intrinsic to the diffusion model to utilize conditional images as a reference. Such integration significantly enhances the model's capability in performing cross-modal image generation, leveraging the contextual information provided by the conditional inputs to inform the generation process. 

ControlNet integrates the architecture of Stable Diffusion with an additional U-Net-based mechanism that utilizes spatial information from reference images as conditional inputs. This integration incorporates a cross-attention mechanism, which significantly enhances the fidelity of the generated images. Both diffusion models, namely Stable Diffusion and the U-Net-based component, are renowned and efficacious in their capacity to synthesize new images by leveraging existing image information.

\begin{figure*}[!t]
\centering
\includegraphics[width=7.16in, height=3.5in]{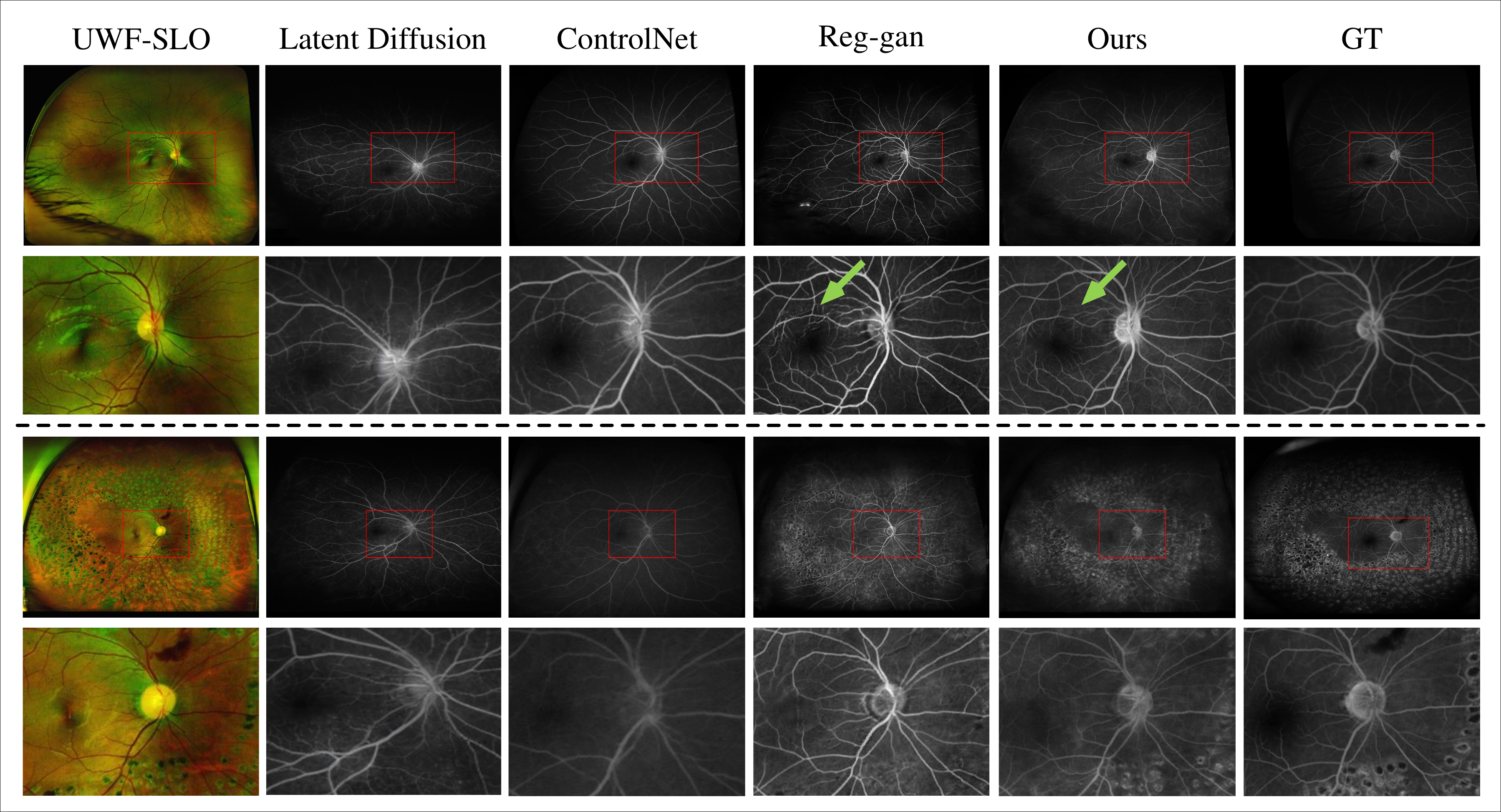}
\caption{Visual comparison of comparative methods generating UWF-FA from UWF-SLO. The ground truth images are depicted in the sixth column, and the bottom row offers magnified views of the corresponding images from the top row. }
\label{zoom-in}
\end{figure*}

\subsubsection{Implementation details}
In our study, we employed the Pix2pix model, a 6-stage U-Net with seven encoders and decoder blocks each, consisting of convolution, normalization, and activation layers. The Pix2pixHD model used a two-tier system combining a 2-stage and a 1-stage U-Net to produce an intermediate image for multiple discriminators.

Reg-GAN's registration network adopted a 7-stage U-Net architecture. The output from this network is fed into a spatial transformer network, as described in \cite{jaderberg2016spatial}, to perform affine transformations. To train the latent and stable diffusion models, a variational autoencoder (VAE) \cite{kusner2017grammar} is initially employed to compress the images into a lower-dimensional latent space. The latent diffusion network utilizes a 5-stage U-Net architecture augmented with skip connections and incorporates self-attention mechanisms at the penultimate layer. Conversely, the stable diffusion network employs a 4-stage U-Net architecture, also featuring skip connections, with self-attention mechanisms applied across the first three layers.

The training of comparative methods strictly followed the protocols outlined in their published source codes to ensure consistency. Nonetheless, due to the well-documented training instability associated with GAN-based networks, our approach adopted a phased optimization strategy. In the initial phase, we confined optimization exclusively to the Fine and Coarse generators. This initial optimization continued for 10 epochs to establish a stable foundation. Subsequently, discriminators were introduced into the training process after the initial 10 epochs, and their assessments were incorporated into the loss function to refine the generative models. After 24 epochs training, we integrated the registration module, at which the correction loss was incorporated as an essential element of the composite loss function. The training persisted until convergence was achieved, which occurred after 40 epochs.

All methods were implemented using the PyTorch 1.12 framework trained on two Nvidia A100 GPUs. We maintained a constant learning rate of 0.0002 throughout the experiments. The Adam optimizer was used with $\beta_1$ and $\beta_2$ parameters set to 0.5 and 0.999, respectively, while other parameters defaulted to the settings provided in the open-source codes published by their authors. For our method, we assigned weights $\lambda_{FMF}$, $\lambda_{FMC}$ with 10 and $\lambda_{VGGF}$, $\lambda_{VGGC}$ with a value of 5.

\subsection{Inception Metrics Study}
The efficacy of our proposed model was benchmarked against the five advanced models previously described, employing four established evaluation metrics: FID$(\downarrow)$, IS$(\uparrow)$, PSNR$(\uparrow)$, and MS-SSIM$(\uparrow)$. The FID is a metric that quantifies the dissimilarity between feature vectors of real and generated images, where these features are extracted utilizing the Inception v3 image classification model. A lower FID score suggests that the classification model experiences greater difficulty in differentiating between real and synthetic images, implying a higher quality of the generated image. In conjunction with the IS, the MS-SSIM, and the PSNR are utilized to evaluate the disparity in the distribution between real and generated images. The IS specifically assesses the quality and diversity of image pairs by calculating the softmax probability distribution over these pairs. The MS-SSIM, on the other hand, is a metric that captures the multi-scale structural similarity between images, taking into account aspects such as luminance, contrast, and structural information. Lastly, PSNR measures the distortion between paired images by calculating the mean squared error, offering insight into the level of degradation or quality loss in the generated images.

In addition, we separately trained these methods on the $Intra$ and $Combined$ datasets constructed in \ref{preparation}, keeping other parameters consistent, to validate the performance of these methods on datasets of different sizes. We also split test datasets from these two datasets, and evaluate the aforementioned metrics on these test datasets.

The outcomes of this comparison are presented in Table. \ref{tab1}. Specifically, in the task of generating UWF-FA images from UWF-SLO in smaller data size, our methodology exhibits a remarkable improvement of 12.075, 0.181, 0.194, and 3.812, compared to the most competitive existing methods in the FID, IS, MS-SSIM and PSNR, respectively. When the amount of data increases, the performance of diffusion-based methods has shown significant improvement. In terms of metrics, our methodology exhibits improvements in FID, IS, and PSNR by 6.505, 0.193, and 2.682, respectively, and only lagged by 0.036 in terms of MS-SSIM.

\begin{figure}[!tbp]
\centerline{\includegraphics[width=\columnwidth]{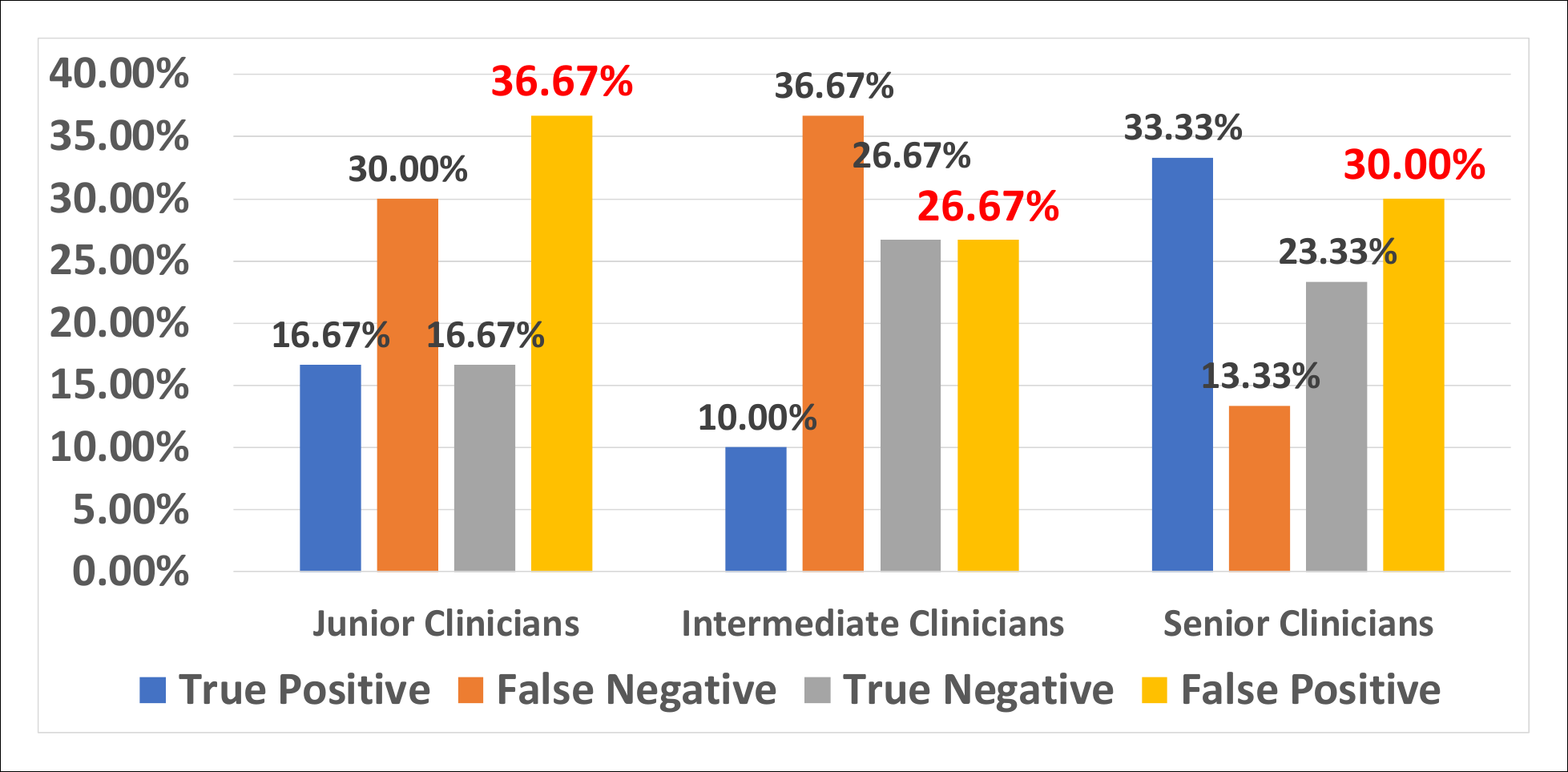}}
\caption{Comparison of the judgment results across clinicians of varying expertise levels. The red digits represent the proportion of generated images considered as authentic one. }
\label{judge}
\end{figure}

\subsection{Visual Comparison}
In Fig. \ref{zoom-in}, we present a visual comparison of various benchmarked methods, illustrating their capability to generate images that capture both local and global features. For this comparison, a UWF-SLO image was randomly selected from the test set, and the outputs generated by different methods were displayed, with zoomed-in views provided for detailed inspection on the next row. Upon comparing these outcomes with the ground truth, illustrated in the rightmost column, it is evident that the images generated by ControlNet and Latent Diffusion exhibit commendable overall quality. This is observed in aspects such as the symmetry of certain vascular structures and the general luminosity of the images. Nevertheless, it is noteworthy that local features, including capillaries and the intricate details of lesions, are not as accurately reproduced. Furthermore, the diffusion-based method demonstrates more significant variances in overall brightness when contrasted with the GAN-based approach. A specific instance of a laser spot transition, presented in rows 3 and 4 of Fig. \ref{zoom-in}, reveals that both ControlNet and Latent Diffusion fall short in achieving the comprehensive lesion transformation observed in Reg-GAN and our proposed method.

The enlarged images on the second row in Fig.  \ref{zoom-in} show that although Reg-GAN's blood vessels appear to be clearer, they are inferior to our method in terms of vessel restoration in the optic disc and optic disc itself. At the same time, Reg-GAN produced false structures (e.g., broken and stuck together blood vessels) in local areas, as indicated by the green arrows. In the case of laser spots, our generated method is closer to the ground truth, transforming almost all laser spot areas and gaining clinical recognition. At the same time, it can be seen from the enlarged images that our method maintains a high level of blood vessel and optic disc generation while completing the transformation of the lesion area.

\subsection{Ablation Study}
In order to generate UWF-FA more realistically and robustly from UWF-SLO, we propose an Image Sharpening (ImS) UWF-SLO pre-processing method in \ref{preparation}, a registration module (RM) in \ref{register}, and two additional loss functions: VGG loss (VGG) and Feature Mapping loss (FM) in \ref{loss_func}. 

To assess the proposed improvement methods' effectiveness, we conducted an ablation study using four Inception metrics to evaluate image quality and experimented with four alternative designs: 1) Omitting the passage of UWF-FA images from the Fine Generator to the RM and, correspondingly, not incorporating the Corr loss; 2) Utilizing the original UWF-SLO images as input without the ImS process and excluding the combined loss function for the Coarse and Fine generators; 3) VGG imaging; and 4) FM imaging. These modified designs, along with our original method, were trained sequentially while maintaining consistency in all other training parameters. We then assessed the performance of each model using the same test dataset across the four established metrics. The outcomes of the five experiments are concisely presented in Table. \ref{ablation_table}.

The impact of four modules or processes on image quality from lowest to highest is ImS, VGG, FM, and RM. Omitting ImS, there is an average decline of 6.77\% in four inceptions and 15.4\% without VGG. Without FM, FID, IS, MS-SSIM and PSNR have deteriorated by 3.598, 0.414, 0.136, 8.8, and 16.3\% on average. In the absence of RM, these metrics have worsened by 8.497, 0.418, 0.365, 12.245, and 28.7\% on average.

% First, our method demonstrates a greater resilience to increasing noise levels compared to Pix2PixHD, regardless of the use of the registration module. This resilience becomes more pronounced as the noise level intensifies. Second, integrating the registration module into UWAFA-GAN consistently enhances the quality of the generated images, with the improvement being particularly notable at lower noise levels (0, 1, and 2). The difference in quality diminishes at higher noise levels.

% Additionally, our method shows an initial improvement in the quality of generated images as noise levels increase, peaking at noise level 2, before declining at higher noise levels. This pattern indicates that a moderate amount of rotation, translation, and scaling can be beneficial, akin to the positive effects of data augmentation during the training phase.

\begin{table}[!t]
\centering
\renewcommand{\arraystretch}{1.4} 
\caption{Ablation study of RM (Registration Module), ImS (Image Sharpening), VGG (VGG Loss) and FM (Feature Mapping Loss) through metrics.}
\centering
    \centering
    \begin{adjustbox}{width=0.49\textwidth, center}
    \resizebox{\linewidth}{!}{%
        \begin{tabular}{c|c|c|c| c c c c}
        \hline
        \multicolumn{4}{c|}{Components} & \multicolumn{4}{c}{Metrics} \\
        \hline
        \multicolumn{1}{c|}{RM} & \multicolumn{1}{c|}{ImS} & \multicolumn{1}{c|}{VGG}& \multicolumn{1}{c|}{FM} & \multicolumn{1}{c}{FID $\downarrow$} & \multicolumn{1}{c}{IS $\uparrow$}& \multicolumn{1}{c}{MS-SSIM $\uparrow$} & \multicolumn{1}{c}{PSNR $\uparrow$} \\
        \hline
         & $\checkmark$&$\checkmark$ & $\checkmark$& 29.923 & 1.254 & 0.568 & 25.466 \\
        $\checkmark$ & &$\checkmark$ & $\checkmark$& 22.391 & 1.488 & 0.872 & 32.73 \\
        $\checkmark$ & $\checkmark$ & & $\checkmark$& 23.773 & 1.386 & 0.929 & 30.553 \\
        $\checkmark$ & $\checkmark$ & $\checkmark$ & & 25.024 & 1.258 & 0.797 & 28.911 \\
        $\checkmark$ & $\checkmark$ & $\checkmark$ & $\checkmark$& \textbf{21.426} & \textbf{1.672} & \textbf{0.933} & \textbf{37.711}\\
        \hline
        \end{tabular}
    }
    \end{adjustbox}
\label{ablation_table}
\end{table}

\subsection{Clinic User Study}
To further assess UWAFA-GAN's clinical value, the synthesized images need not only meet technical standards but also appear genuine to expert ophthalmologists. In this assessment, we involved ophthalmologists at three expertise levels, senior (10 years of experience, Attending Doctors), intermediate (5 years, Intermediate Physicians), and junior (3 years, Resident Physicians), in evaluating fluorescein angiography.

These ophthalmologists were tasked with evaluating a specially curated dataset, which was constructed as follows: a random selection of UWF-SLOs and their true UWF-FA counterparts were chosen, to which our model-generated UWF-FAs were added, creating sets of image pairs. Each pair consisted of a UWF-SLO and either a true UWF-FA or a synthetic UWF-FA, selected at random. A total of 30 such pairs were compiled for the evaluation. Each evaluator was presented with the 30 image pairs and instructed to determine whether the UWF-FAs were genuine or synthesized by UWAFA-GAN, using the corresponding UWF-SLOs as a reference. Ophthalmologists were asked to assess each pair of images within 30 seconds, based on the successful conversion of lesions on the UWF-FA and the overall image fidelity.

Upon completion of the exercise, the clinicians' judgments were aggregated, compared with the correct classifications, and tallied into correct and incorrect responses. Correct responses were then subdivided into true positives and true negatives, while incorrect responses were parsed into false negatives and false positives. We then calculated the percentages for each category and averaged the outcomes for clinicians at the same level of expertise. Fig. \ref{judge} presents these results in histogram format.

\begin{figure}[!tbp]
\centerline{\includegraphics[width=\columnwidth]{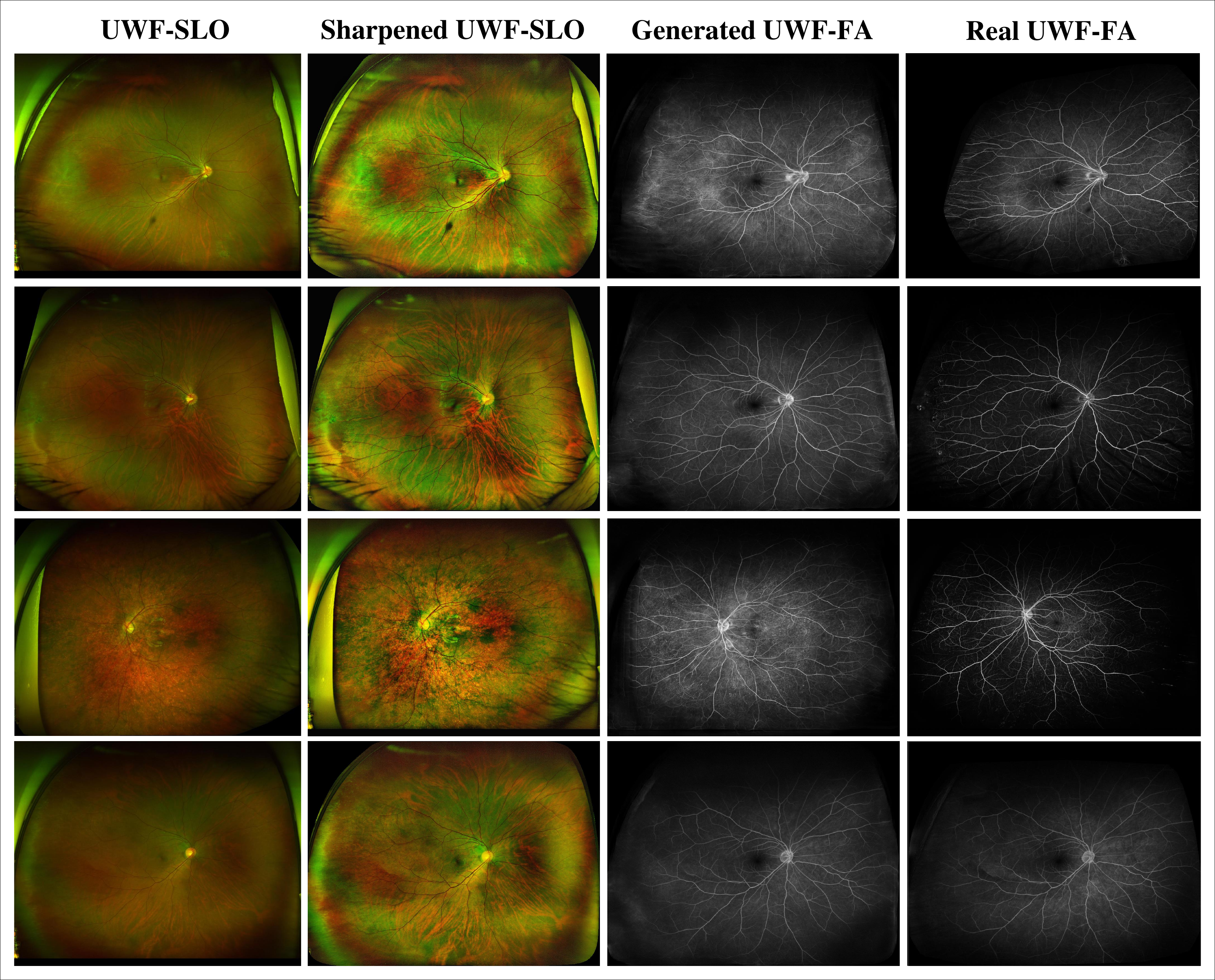}}
\caption{Four pairs of UWF images that elicited false positives in our clinical user study.}
\label{good}
\end{figure}

The histogram indicates that junior clinicians exhibited the highest rate of false positives, misclassifying 36.67\% of the generated UWF-FAs as authentic. Intermediate and senior clinicians also showed notable false positive rates of 26.67\% and 30\%, respectively. When examining only the synthetic UWF-FAs, the percentages identified as real by junior, intermediate, and senior clinicians were 68.75\%, 50\%, and 56.25\%, respectively. These findings suggest that at least half of the model-generated UWF-FAs were convincingly realistic to the evaluators. To visually underscore these observations, we selected four image pairs that elicited false positive judgments across all levels of expertise for display in Fig. \ref{good}. These pairs serve as prime examples of the convincing and high-quality images produced by our method.

\begin{table}[!t]
\caption{The Acc, F1 score, and TL of ResNet101 and SeNet101 with/without generated UWF-FA by our method. }
\resizebox{\columnwidth}{!}{%
\begin{tabular}{@{}cccc|ccc@{}}
\toprule
\multicolumn{1}{c|}{Methods} & \multicolumn{3}{c|}{ResNet101 \cite{he2016deep}}                   & \multicolumn{3}{c}{SeNet101 \cite{hu2018squeeze}}                     \\ \midrule
\multicolumn{1}{c|}{Metrics} & Acc.           & F1             & TL             & Acc.           & F1             & TL             \\ \midrule
\textit{w/o UWF-FA}          & 76.19          & 0.818          & 0.563          & 79.76          & 0.841          & 0.438          \\
\textit{w UWF-FA}            & \textbf{80.95} & \textbf{0.857} & \textbf{0.347} & \textbf{82.14} & \textbf{0.859} & \textbf{0.387} \\ \bottomrule
\end{tabular}%
}
\label{downstream}
\end{table}

\begin{figure}[!t]
\centerline{\includegraphics[width=\columnwidth, height=2in]{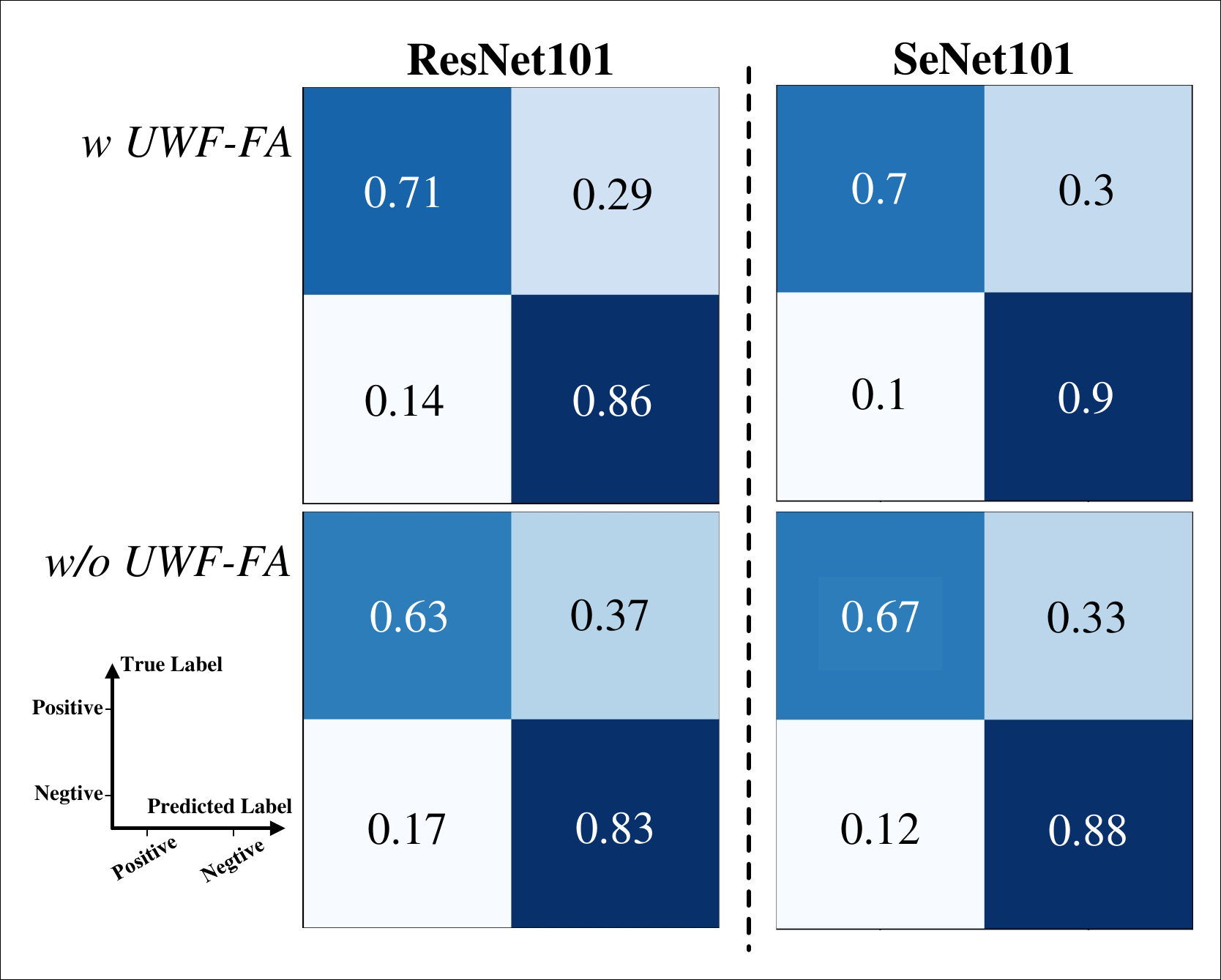}}
\caption{Confusion matrices of evaluation results of ResNet101 and SeNet101 with and without generated UWF-FA. }
\label{confusion}
\end{figure}

\subsection{Downstream Task Study}

To showcase the enhanced performance of our method on downstream tasks, we constructed a small binary classification dataset to validate our model's superior capability in such tasks, utilizing standard classification techniques. A group of ophthalmologists was enlisted to label 464 UWF-SLO images as either lesion-present or lesion-absent, yielding a training set distribution of 258:122 and a test set ratio of 47:37.

For the classification task, we selected two of the most reputable algorithms: ResNet101\cite{he2016deep} and SeNet101\cite{hu2018squeeze}. We conducted two sets of experiments with each classification approach: 1. Utilizing only the UWF-SLO images as input, referred to as \textit{w/o UWF-FA}; 2. Combining the UWF-SLO images, processed through our generative model to produce corresponding UWF-FA images, with the original UWF-FA images along the channel dimension prior to input into the classification model, denoted as \textit{w UWF-FA}. All experiments were trained over 50 epochs, with parameters from the best-performing epoch retained. Subsequently, we evaluated the Accuracy (Acc.), F1 score (F1), and Testing Loss (TL), and recorded the corresponding confusion matrix on the test set for all experimental setups.

According to the results summarized in Table.  \ref{downstream}, the ResNet101 classification model demonstrates significantly enhanced performance when processing both UWF-SLO and UWF-FA inputs simultaneously, showcasing improvements of 4.76\% in Accuracy, 0.039 in F1 score, and a decrease of 0.216 in Training Loss compared to using only UWF-SLO inputs. Similarly, the SeNet101 model records improvements of 2.38\% in Accuracy, 0.018 in F1 score, and a decrease of 0.051 in Training Loss. Moreover, the confusion matrix illustrated in Fig. \ref{confusion} indicates that integrating synthesized UWF-FA information not only augments the model's precision in identifying negative samples but also elevates the accuracy of positive sample identification. This underscores the added value of our model in providing additional lesion information, thus empowering the classification algorithm to render more accurate determinations.

\section{Conclusion}
To address the potential risks associated with the administration of fluorescein dye in UWF-FA, we introduce UWAFA-GAN engineered to synthesize UWF-FA images from UWF-SLO images. Our method is adept at producing high-resolution outputs that capture micro-vascular anomalies and lesions while demonstrating robustness against noise in the training datasets. As evidenced by its performance on a proprietary dataset, UWAFA-GAN surpasses existing methods in terms of clinical applicability and provides pathological changes in high resolution and image quality compared to previous synthesis techniques.

% Despite its strengths, our method is not without limitations. First, the accuracy of our generated images, particularly in regions with exceedingly small lesions, is not always optimal; certain fine lesions and the optic disc may be rendered with less clarity. Second, the limited size of our current dataset restricts the model's ability to learn from a broader spectrum of rare lesion types, which could contribute to its sub-optimal performance in some areas.

Despite the commendable attributes of our methodology, it is not devoid of constraints. Firstly, within the UWF-FA, there exist certain rare lesions that have an impact on physicians' diagnoses. However, due to their sparse distribution in the overall data, our method does not achieve high accuracy in generating these rare lesions. Secondly, a more intimate connection can be established between the generation framework and downstream tasks. For instance, if our approach could accurately generate while simultaneously segmenting lesions, it would hold greater clinical value. 

To address these challenges, future work will concentrate on explicitly incorporating information about the pathology. We also plan to expand the dataset, specifically by incorporating more instances of rare lesion types, to facilitate more comprehensive learning by the model. Our objective is to further enhance and validate this technology with the ambition of establishing UWAFA-GAN as a dependable supplementary tool for clinical diagnosis and the detection of fundus diseases. 

\section{Compliance With Ethical Standards}
This research study was conducted retrospectively and following the principles of the Declaration of Helsinki. Our study included an informed consent waiver and was approved by the Ethics Committee of Shenzhen Eye Hospital (2022KYPJ073). 
\section*{References}
\vspace{-2em}
\bibliographystyle{IEEEtran}
\bibliography{main}
\end{document}